\documentclass[pre]{revtex4}
\usepackage{amsmath,amssymb,latexsym,epsfig,graphics,epsf,bm}
\usepackage{graphics}
\usepackage{float}
\usepackage{placeins}
\newcommand{\br}{\mathbf r}
\newcommand{\bV}{\mathbf V}

\newcommand{\bR}{\mathbf R}

\newcommand{\tti}{\xi}

\newcommand{\be}{\begin{equation}}
\newcommand{\ee}{\end{equation}}

\renewcommand{\o}{\char 28}
\usepackage{color}
\begin{document}

\title{Narayanaswamy's 1971 aging theory and material time}
\author{Jeppe C. Dyre}\email{dyre@ruc.dk}
\affiliation{DNRF Center ``Glass and Time", IMFUFA, Department of Sciences, Roskilde University, Postbox 260, DK-4000 Roskilde, Denmark}
\date{\today}

\begin{abstract}
The Bochkov-Kuzovlev nonlinear fluctuation-dissipation theorem is used to derive Narayanaswamy's phenomenological theory of physical aging, in which this highly nonlinear phenomenon is described by a linear material-time convolution integral. A characteristic property of the Narayanaswamy aging description is material-time translational invariance, which is here taken as the basic assumption of the derivation. It is shown that only one possible definition of the material time obeys this invariance, namely the square of the distance travelled from a configuration of the system far back in time. The paper concludes with suggestions for computer simulations that test for consequences of material-time translational invariance. One of these is the ``unique-triangles property'' according to which any three points on the system's path form a triangle such that two side lengths determine the third; this is equivalent to the well-known triangular relation for time-autocorrelation functions of aging spin glasses [L. F. Cugliandolo and J. Kurchan, J. Phys. A: Math. Gen. {\bf 27}, 5749 (1994)]. The unique-triangles property implies a simple geometric interpretation of out-of-equilibrium time-autocorrelation functions, which extends to aging a previously proposed framework for such functions in equilibrium [J. C. Dyre, cond-mat/9712222].
\end{abstract}

\maketitle

\section{Introduction}\label{intro}

Physical aging is the small, gradual change of material properties observed for some systems, e.g., polymers and glasses just below the glass-transition temperature \cite{scherer}. In contrast to the aging of most real-life materials resulting from chemical reactions, physical aging is exclusively due to adjustments of molecular positions. During the last 50 years there has been slow, but steady progress in the description and understanding of aging in experiments as well as model systems. For example, studies of spin glass aging in the 1990s led to the concept of an ``effective'' temperature that quantifies violations of the fluctuation-dissipation theorem describing linear response around a state of thermal equilibrium \cite{cug94,cug97}. There is still no good theoretical justification, however, of Narayanaswamy's remarkable phenomenological physical-aging theory from 1971 \cite{nar71}, which has been used routinely in industry for decades \cite{scherer}. 

This paper is an attempt to understand the microscopic origin of Narayanaswamy's theory in which physical aging, which is strongly {\it non-linear} in the temperature history, is described by a {\it linear} convolution integral if one replaces the time integration variable by the so-called material time. The material time quantifies how fast processes take place in an aging system, reflecting the existence of an ``inner clock''. In the words of Narayanaswamy's seminal paper \cite{nar71}: ``The intrinsic relation between force and flow (or cause and effect) is assumed to be linear, i.e., it is assumed that the only cause of nonlinearity in glass transition phenomena is the changing viscosity of the glass as the structure (or fictive temperature) changes.''

The Narayanaswamy theory, which works best for small temperature variations \cite{scherer}, raises a number of questions:

\begin{enumerate}
\item  How and why can the highly non-linear physical-aging phenomenon be described by a linear convolution integral?
\item What is the material time? Can it be defined in terms of the system's microscopic variables? 
\item Are all aging quantities subject to the same material time? 
\item Which physical quantity controls the material time?
\end{enumerate}

The first three questions are addressed in this paper, which derives the Narayanaswamy theory from an exact, non-linear fluctuation-dissipation theorem, assuming material-time translational invariance (Sec. \ref{nara}). The microscopic expression for the material time is derived in Sec. \ref{clock}. Sections \ref{conseq} and \ref{tests} discuss consequences and give a few suggestions for numerical tests of the ideas developed; Sec. \ref{end} concludes the paper.

We do not claim originality of any arguments and results of this paper, which is just an attempt to illuminate the Narayanaswamy 1971 aging theory by connecting it to the 1981 Bochkov-Kuzovlev non-linear fluctuation-dissipation theorem and the 1994 spin-glass aging theory of Cugliandolo and Kurchan \cite{cug94}. -- Much has happened in physics and chemistry since 1971, but apparently the theory of aging itself matures only like a slowly aging system.

\section{Background}\label{back}

This section summarizes some necessary background.

\subsection{The material-time concept and Narayanaswamy's phenomenological aging theory}

Physical aging is usually studied in glasses and glass-forming liquids. For these systems the relaxation time is dramatically temperature dependent (sometimes increasing up to a factor of ten for a temperature decrease of just one percent). Even the simplest case, that of a temperature jump from a state of equilibrium of a highly viscous liquid, is {\it nonlinear} unless the temperature change is very small \cite{scherer}. Another complication is that aging is virtually always {\it nonexponential} in time \cite{scherer,dyr06}. 

Interest in physical aging has throughout the years been stimulated by the practical importance of being able to predict aging quantitatively, and there is a large literature on physical aging experiments \cite{kov63,moy76,str78,scherer,hut95,mck95,mau09a,can13,zha13}. An excellent and still not outdated introduction to the phenomenology of aging is given in Scherer's book from 1986 \cite{scherer}. Experiments have studied physical aging by probing, e.g., density \cite{kov63,sim97}, enthalpy \cite{nar71,moy76}, Young's modulus \cite{che78}, gas permeability \cite{hua04}, high-frequency mechanical moduli \cite{ols98,dil04}, dc conductivity \cite{str78}, frequency-dependent dielectric constant \cite{leh98,lun05,hec15}, beta relaxation loss-peak frequency \cite{hec15}, X-ray photon-correlation-spectroscopy probed structure \cite{rut12}, non-linear dielectric susceptibility \cite{bru12}, etc. In some cases one monitors aging following a temperature jump from a state of equilibrium, but more complicated temperature histories have also been studied extensively, in particular that of continuous cooling of a liquid into its glass state. 

To be specific we shall assume that the property $X$ is monitored when temperature is changed in a controlled fashion, and that the experiment ends in equilibrium at the temperature $T_0$. The deviation of $X$ from its equilibrium value at $T_0$ is denoted by $\Delta X(t)$; thus $\Delta X(t)\rightarrow 0$ as $t\rightarrow\infty$. Narayanaswamy's idea, which was inspired by rheological models, was to introduce a so-called material time, $\tti$ \cite{nar71}. This quantity may be thought of as the time measured on a clock with a clock rate that {\it itself ages} \cite{nar71,hec10}, reflecting the fact that processes in an aging system proceed with a speed that changes as the structure ages. Following a down jump in temperature, for instance, the clock rate decreases gradually; it eventually becomes constant, but only when the system is very close to equilibrium. The material time is intuitively analogous to the proper time of relativity theory, which is the time measured on a clock moving with the observer.

If the clock rate is denoted by $\gamma(t)$, the material time $\xi(t)$ is defined by 

\be\label{eq1}
d\tti=\gamma(t)\,dt\,.
\ee
In thermal equilibrium the clock rate is constant and the material time is proportional to the laboratory time, though with a proportionality constant $K$ that is highly temperature dependent, i.e., $\xi_{\rm eq}(t)=K(T)t+{\rm Const.}$ For an aging system, the clock rate is generally a complicated function of the system's thermal history after it fell out of equilibrium. 

Narayanaswamy suggested that for the temperature variation $T(t)=T_0+\Delta T(t)$, if rewritten as a function of $\xi$ the variation $\Delta X(t)$ has an ``instantaneous'' contribution, $C\Delta T(t)=C\Delta T(\xi(t))$, and a relaxing contribution given by a {\it linear} material-time convolution integral \cite{nar71,scherer}, i.e.,

\be\label{Nar}
\Delta X(\tti_2)=C\Delta T(\tti_2)-\int_{-\infty}^{\tti_2} M_X(\tti_2-\tti_1)\, \frac{d\Delta T}{d\tti_1}(\tti_1)\,d\tti_1\,.
\ee
As an example, after a temperature jump at $t=0$ from equilibrium at temperature $T_0+\Delta T$ to the temperature $T_0$ one has $\Delta X(t)=\Delta T M_X(\tti(t))$ in which $\tti(t)$ is found by integrating Eq. (\ref{eq1}). The crucial point is that the function $M_X(\tti)$ is the same for all temperature jumps whereas the material time $\xi(t)$ depends on $\Delta T$ (usually strongly). 

The Narayanaswamy formalism represented a great step forward compared to that time's use of non-linear differential equations for describing aging. Such equations cannot reproduce memory effects like the crossover (Kovacs) effect \cite{scherer}, and they do not reflect the deep insight that aging can be described by a {\it linear} theory. Recently, the use of differential equations for describing aging has again been investigated, but now limited to temperature jumps. Thus Kolvin and Bouchbinder have proposed a simple temperature-jump differential equation that fits data well \cite{kol12}, and Hecksher {\it et al.} showed that for aging following temperature jumps the Narayanaswamy formalism is equivalent to a differential equation \cite{hec15}. The latter differential equation involves both the normalized and the unnormalized relaxation function, and this is necessary to ensure consistency with Eq. (\ref{Nar}). This equation results in a simple expression for calculating one nonlinear relaxation function from another, without any fitting to analytical functions as is usually done in applications of the Narayanaswamy formalism. The resulting formalism was confirmed in experiments monitoring dielectric relaxation and the high-frequency shear modulus during aging \cite{hec15}.

In the Narayanaswamy theory the clock rate of the aging system's ``inner clock'' is a global variable. An interesting extension of this formalism is due to Castillo and co-workers \cite{cas07,par09,avi11} who assume that aging is controlled by a space-dependent clock rate. This idea, which makes perfect sense in view of the existence of dynamical heterogeneities in glass-forming liquids and glasses \cite{boh96,vol02}, has been verified numerically for different model systems \cite{cas07,par09,avi11}. Nevertheless, below we do not follow this promising idea, but focus on the simpler case of a ``global average'' clock.

\subsection{Material-time translational invariance (MTTI)}

Equation (\ref{Nar}) has the appearance of standard linear-response theory, though with the time $t$ replaced by the material time $\xi$ . A characteristic property of linear-response theory is time-translational invariance, which expresses the fact fundamental to all science that repeating an experiment at a later time leads to the same result. An implicit assumption of Narayanaswamy's formalism is that translational invariance applies for the material time during aging as reflected in the fact the memory function in Eq. (\ref{Nar}) is a function of the {\it difference} $\xi_2-\xi_1$. This will be referred to as material-time translational invariance (MTTI). 

Translated into standard statistical-mechanical language, MTTI is the property that even during aging, time correlation functions depend only on the difference of the material times involved, i.e., that for any two quantities $A$ and $B$ a function $F_{AB}(\xi)$ exists such that

\be\label{MTTdef}
\langle A(t_1)B(t_2)\rangle
\,=\, F_{AB}\big(\xi(t_2)-\xi(t_1)\big)\,.
\ee
The angular brackets denote an ensemble average (for the aging system). In contrast to the equilibrium situation, the left-hand side is generally {\it not} a function of $t_2-t_1$. It is understood here and henceforth that $A$ and $B$ both have zero average -- if that is not the case, the average value should be subtracted at each time $t$.

Of particular interest is MTTI for the displacement vector that specifies the system's path in configuration space, the $3N$-dimensional vector $\bR\equiv (\br_1,...\br_N)$ where $\br_i$ is the position of the $i$th particle. One way to investigate a system's dynamics is to evaluate the mean-square displacement from one time $t_1$ to a later time $t_2$, $\langle\big(\bR(t_2)-\bR(t_1)\big)^2\rangle$. This quantity, of course, refers to the total displacement, which implies going beyond the boundaries of standard computer simulations performed with periodic boundary conditions \cite{sad13,de14}. With this clarification in mind, MTTI for the mean-square displacement means that it depends only on the difference in material times involved, i.e., 

\be\label{MTTR2}
\langle \big(\bR(t_2)-\bR(t_1)\big)^2\rangle
\,=\, N\,G\big(\xi(t_2)-\xi(t_1)\big)\,.
\ee

\subsection{The triangular relation}\label{triangsec}

More than 20 years ago the study of aging in spin glasses resulted in a deep insight into the nature of aging in such discrete models \cite{cug94}. Interestingly, this work seems to have been carried out with little interaction with the experimental oxide glass and polymer aging communities, which  had developed over long time \cite{nar71,moy76,str78,hut95}. 

For spin glass aging there is time-scale separation, just as for the aging of ordinary glasses. On short time scales an analog of vibrational thermal equilibrium exists, whereas aging of a spin glass' structural degrees of freedom takes place on much longer time scales. This led to the introduction of the concept of an {\it effective temperature} \cite{cug94,cug97,cha07} that is conceptually similar to Tool's intuitive {\it fictive temperature} concept from 1946 \cite{too46}, but founded in rigorous statistical-mechanical theory. 

Moreover, the spin glass works resulted in the understanding that "the proper measure of time is the correlation itself, not the laboratory clock'' \cite{cha07}. This was formalized into the {\it triangular relation} which, if $C(t,t')$ is a time-autocorrelation function, i.e., an average for the aging system of the form  $C(t,t')\equiv\langle A(t)A(t')\rangle$, is the following equation valid on long time scales ($t_1<t_2<t_3$) \cite{cug94}

\be\label{triang}
C(t_1,t_3)
\,=\, f\big(C(t_1,t_2),C(t_2,t_3)\big)\,.
\ee
The idea that time is measured via the autocorrelation function itself \cite{cug94,cha07} implies time-reparametrization invariance, see, e.g., Refs. \onlinecite{cas01,cha02,avi11}, as well as the earlier Refs. \onlinecite{som81} and \onlinecite{hor84}. 

As detailed shortly, the triangular relation is equivalent to MTTI, but first a few notes:

\begin{enumerate}

\item Equation (\ref{triang}) applies in thermal equilibrium for any time-autocorrelation function that is a sum of decaying exponentials. In that case, the value of the time autocorrelation function $C(t_1,t_2)$ uniquely determines the time difference $t_2-t_1$, and $C(t_2,t_3)$ likewise determines $t_3-t_2$, which means that knowledge of both $C(t_1,t_2)$ and $C(t_2,t_3)$ implies that $t_3-t_1=(t_3-t_2)+(t_2-t_1)$ and thus $C(t_1,t_3)$ are known, which is the triangular relation Eq. (\ref{triang}).

\item For the case of equilibrium, if time-temperature superposition (TTS) applies, the function $f$ is the same when temperature is changed (for the normalized autocorrelation functions). Conversely, if Eq. (\ref{triang}) applies for aging over a certain range of temperatures, this implies TTS.

\item The triangular relation applies whenever the system in question obeys so-called dynamic ultrametricity, i.e., when $C(t_1,t_3)={\rm min}\left(C(t_1,t_2),C(t_2,t_3)\right)$. This condition  is believed to describe the aging of most or all spin glasses \cite{fra00}.

\end{enumerate}

The triangular relation applies for any aging system that has an ``inner clock'' and a corresponding material time $\xi$ obeying MTTI. In that case, if the function $F_{CC}(\xi)$ of Eq. (\ref{MTTdef}) goes monotonically to zero as $\xi$ goes to infinity, $C(t_1,t_2)$ determines the difference $\xi(t_2)-\xi(t_1)$. Likewise, $C(t_2,t_3)$ determines the difference $\xi(t_3)-\xi(t_2)$. Thus $C(t_1,t_2)$ and $C(t_2,t_3)$ determine $C(t_1,t_3)$, which is the triangular relation. Note the analogy to the equilibrium argument.

Conversely, if a system ages in such a way that its time-autocorrelation functions obey the triangular relation, one can define a material time $\xi(t)$ that satisfies MTTI. This nontrivial result was derived in Appendix B of Ref. \onlinecite{cug94}. We will not here repeat the derivation involving formal group theory, but briefly the argument is the following. Assuming again that $C(t_1,t_2)$ goes monotonically to zero as $|t_2-t_1|\rightarrow\infty$, one defines an algebraic operation $*$ on the part of the positive real axis that corresponds to some value of the autocorrelation function: For $x,y,z$ in this part of the real axis, we define $x*y\equiv f(x,y)$ where $f$ is the function appearing in the triangular relation Eq. (\ref{triang}). It is straightforward to show from Eq. (\ref{triang}) that $*$ is associative, i.e., that $x*(y*z)=(x*y)*z$. The commutative law also applies, i.e., $x*y=y*x$, which is not trivial -- this signals an intriguing element of time reversibility during aging. From these facts one can show that functions $\tilde h(t)$ and $\phi(\tilde h)$ exist such that $C(t_1,t_2)=\phi\big(\tilde h(t_2)-\tilde h(t_1)\big )$ \cite{cug94}. This is Eq. (\ref{MTTdef}) with $\tilde h(t)=\xi(t)$, i.e., the triangular relation leads to the existence of a material time and implies MTTI.

\section{Deriving Narayanaswamy's aging theory from MTTI}\label{nara}

The fundamental puzzle of the Narayanaswamy formalism is how to arrive at a theory that describes aging by a {\it linear} material-time convolution integral. To show how this may come about we combine MTTI with Bochkov's and Kuzovlev's non-linear fluctuation-dissipation theorem from 1981 \cite{boc81}.

It is convenient to adopt the language of the energy-bond formalism according to which any system interacts with its surroundings via one or more so-called energy bonds \cite{paynter,ost71,pvc78,systemdyn,III}. An energy bond is characterized by two variables, an effort and a flow variable, the product of which gives the energy per unit time transferred into the system from its surroundings. Effort could be, e.g., voltage drop or force, in which cases the corresponding flow variables are, respectively, electrical current and velocity. The formalism is general, and even a heat flow may be modelled -- in that case energy is generalized into free energy, the effort is the temperature deviation from a reference temperature $T_0$, and the flow is the entropy current \cite{meixner,bat79}. The general nature of the energy-bond formalism means that for any physical quantity $q$ one may define a flow $f$ from $f\equiv dq/dt$ and let $e$ be a fictitious field coupling linearly to $q$ in the Hamiltonian. The thermal average effort and flow are assumed to be zero.

We first consider the case of a single energy bond in which $e(t)$ is the effort and $f(t)$ the flow. If the effort is externally controlled, the standard linear fluctuation-dissipation (FD) theorem is the following expression for the average flow at time $t_2$ in the time-dependent external field $e(t)$

\be\label{fd1}
\langle f(t_2)\rangle
=\beta_0\int_{-\infty}^{t_2} \langle f(t_2)f(t_1)\rangle_0\, e(t_1)\,dt_1\,.
\ee
Here $\beta_0\equiv 1/k_BT_0$ where $T_0$ is the system's temperature and $\langle f(t_2)f(t_1)\rangle_0$ is the {\it thermal equilibrium} time-autocorrelation function, which by standard time-translational invariance is a function of the difference $t_2-t_1$. There is complete symmetry between $e$ and $f$ in the energy-bond formalism, and if the flow $f$ is externally controlled the FD theorem is

\be\label{fd2}
\langle e(t_2)\rangle
=\beta_0\int_{-\infty}^{t_2} \langle e(t_2)e(t_1)\rangle_0\, f(t_1)\,dt_1\,.
\ee

In 1981 Bochkov and Kuzovlev derived a general, non-linear response theory \cite{boc81}. It is assumed that the system was in equilibrium at the temperature $T_0$ far back in time and ends up in equilibrium at the same temperature in the distant future. If {\it non-equilibrium} cumulant averages are denoted by $\langle ... \rangle_c$, the following exact relation (Eq. (2.18) in Ref. \onlinecite{boc81} in the energy-bond notation) applies:

\be\label{nlfd1}
\langle f(t)\rangle
=\sum_{k=1}^\infty \frac{(-1)^{(k-1)}\beta_0^{k}}{k!}
\int_{-\infty}^{t}dt_1...\int_{-\infty}^{t}dt_k\, \langle f(t)f(t_1)...f(t_k)\rangle_c\, e(t_1).... e(t_k)\,.
\ee
To first order in $e$ all but the first term on the right-hand side may be ignored and the two-point cumulant average may be replaced by the equilibrium product average $\langle f(t_2)f(t_1)\rangle_0$, thus reducing Eq. (\ref{nlfd1}) to the standard FD theorem Eq. (\ref{fd1}). 

The derivation of Eq. (\ref{nlfd1}) is based on Bochkov and Kuzovlev's theorem \cite{boc81} according to which the ratio of probabilities of a given path in time $\bR(t)$ and the corresponding time-reversed path $\bR(-t)$ is the ratio of the Boltzmann probabilities of the starting and ending points of the path (considered over some finite, large time interval). During the 1990s this and related results became known as the fluctuation theorem (see, e.g., Refs. \onlinecite{kur98,cro99,eva02,esp09}); to the best of my knowledge its first general derivation was given in Refs. \onlinecite{boc81} and \onlinecite{boc81a}, covering both the classical Newtonian, stochastic, and quantum-mechanical cases. 

Equation (\ref{nlfd1}) is not the simple higher-order generalization of the linear FD theorem in which the right-hand-side averages are equilibrium averages. In fact, as shown some time ago by Stratonovich \cite{str70}, there is no general nonlinear FD theorem determining the nonlinear response from equilibrium averages \cite{boc81}. The occurrence of {\it non-equilibrium} averages on the right-hand side makes Eq. (\ref{nlfd1}) appear to be fairly useless. We proceed to show how one can, however, based on MTTI use Eq. (\ref{nlfd1}) to arrive at the Naryanaswamy expression that is linear in the material time.

The starting point is the already mentioned fact that Naryanaswamy's theory works best for small temperature variations \cite{scherer}. For larger jumps it breaks down; it is not known whether this is because multiple relaxation mechanisms must be taken into account, or whether there is a fundamental break down of the material-time idea (see, e.g., Ref. \onlinecite{can13a} and references therein). In any case, it is known from experiments that there is an aging regime described by Eq. (\ref{Nar}) in which the temperature changes involved are fairly small (typically a few percent), but large enough that aging is strongly non-linear (e.g., with changes of a factor of ten or more in the characteristic relaxation time between up and down jumps to the same temperature \cite{scherer,mck95,can13,hec15}). 

Standard aging experiments are controlled by temperature, in which cases the effort $e$ is the temperature difference to a reference temperature $T_0$. For the regime of relatively small temperature variations it makes good sense to keep only the lowest-order term on the right-hand side of Eq. (\ref{nlfd1}). Unfortunately, as in most of physics, it is not possible to predict from Taylor expansions how good is the approximation of just keeping the first order term. This means that we cannot estimate how small the temperature jump must be for thisapproximation to be good. 

Ignoring the higher-order terms of Eq. (\ref{nlfd1}) leads to

\be\label{nlfd1a}
\langle f(t_2)\rangle
\cong\beta_0
\int_{-\infty}^{t_2} \langle f(t_2)f(t_1)\rangle_c e(t_1)dt_1\,.
\ee
We have ignored the fact that the cumulant averages of Eqs. (\ref{nlfd1}) and (\ref{nlfd1a}) are themselves in general highly nonlinear functions of the effort variable $e$. The procedure of ignoring this is justified below {\it a posteriori} from MTTI and a change of integration variable to the material time. To consistently ignore higher-order terms one must also ignore the difference between the cumulant average $ \langle f(t_2)f(t_1)\rangle_c$ and the product average $\langle f(t_2)f(t_1)\rangle$ (the difference is $\langle f(t_2)\rangle\langle f(t_1)\rangle$ and of second order in $e$ if $\langle f(t)\rangle_0=0$). This leads to 

\be\label{nlfd1b}
\langle f(t_2)\rangle
\cong\beta_0
\int_{-\infty}^{t_2} \langle f(t_2)f(t_1)\rangle e(t_1)dt_1\,.
\ee
Equation (\ref{nlfd1b}) has the appearance of the standard linear FD theorem Eq. (\ref{fd1}), but since $\langle f(t_2)f(t_1)\rangle$ is evaluated along {\it the actual system path}, this quantity is as mentioned in general a highly nonlinear function of the effort history $e(t)$ and totally different from any equilibrium time-autocorrelation function. The point of MTTI used below is that changing the integration variable to the material time eliminates this effort history dependence.

We rewrite the flow variable $f$ as a time derivative of a generalized ``charge'' $q$, 

\be
f\equiv \frac{dq}{dt}\,
\ee
and change the integration variable in Eq. (\ref{nlfd1b}) from the time $t$ to the material time $\xi=\xi(t)$. This leads to

\be\label{nlfd2}
\left\langle \frac{dq}{d\xi_2}(\xi_2) \right\rangle
\cong\beta_0\int_{-\infty}^{\xi_2} \left\langle\frac{dq}{d\xi_2}(\xi_2)\frac{dq}{d\xi_1}(\xi_1) \right\rangle\, e(\xi_1)\,d\xi_1\,.
\ee
By MTTI (Eq. (\ref{MTTdef})) the autocorrelation function $\langle q(\xi_2){q}(\xi_1)\rangle$ is a function of $\xi_2-\xi_1$, i.e.,

\be\label{ident}
\langle q(\xi_1)q(\xi_2)\rangle
\,=\, F_{qq}(\xi_2-\xi_1)\,.
\ee
The crucial point is that all dependence on $e(t)$ on the left-hand side comes from the material time's effort dependence, whereas the function $F_{qq}$ is independent of $e(t)$.

Equation (\ref{ident}) implies 

\be\label{autcor}
\left\langle\frac{dq}{d\xi_2}(\xi_2)\frac{dq}{d\xi_1}(\xi_1) \right\rangle
\,=\, -F_{qq}''(\xi_2-\xi_1)\,,
\ee
and Eq. (\ref{nlfd2}) thus becomes  

\be\label{nlfd2a}
\frac{d}{d\xi_2}
\langle q(\xi_2) \rangle
\cong -\beta_0\int_{-\infty}^{\xi_2} F_{qq}''(\xi_2-\xi_1)\, e(\xi_1)\,d\xi_1\,.
\ee
Replacing $\xi_2$ with $\xi$ and integrating with respect to $\xi$ from $-\infty$ to $\xi_2$ leads to (with $\Delta q(\xi_2)\equiv q(\xi_2)-q_{\rm eq}(T_0)$)

\be\label{nlfd2b}
\left\langle \Delta q(\xi_2) \right\rangle
\cong  -\beta_0\int_{-\infty}^{\xi_2} d\xi \int_{-\infty}^{\xi} F_{qq}''(\xi-\xi_1)\, e(\xi_1)d{\xi_1}\,.
\ee
Interchanging the order of integration we get since $-\infty<\xi_1<\xi<\xi_2$

\be\label{nlfd2b'}
\left\langle \Delta q(\xi_2) \right\rangle
\cong  -\beta_0\int_{-\infty}^{\xi_2} d{\xi_1}\,  e(\xi_1)\int_{\xi_1}^{\xi_2} \, F_{qq}''(\xi-\xi_1)d{\xi}\,.
\ee
Utilizing that $F_{qq}'(0)=0$, which follows from the fact that $F(\xi)$ also describes the equilibrium linear response for which $(d/dt)\langle q(0)q(t)\rangle |_{t=0}=0$ because of time-reversal invariance,  we get 

\be\label{nlfd2c}
\left\langle \Delta q(\xi_2) \right\rangle
\cong -\beta_0\int_{-\infty}^{\xi_2} F_{qq}'(\xi_2-\xi_1)\, e(\xi_1)\,d\xi_1
= \beta_0\int_{-\infty}^{\xi_2} \left(\frac{d}{d\xi_1} F_{qq}(\xi_2-\xi_1)\right)\, e(\xi_1)\,d\xi_1\,.
\ee
When integrated partially, since $F_{qq}(\xi)\rightarrow 0$ for $\xi\rightarrow\infty$, this leads to

\be\label{nlfd2d}
\left\langle \Delta q(\xi_2) \right\rangle
\cong  \beta_0\left(F_{qq}(0)e(\xi_2)-\int_{-\infty}^{\xi_2}  F_{qq}(\xi_2-\xi_1)\, \frac{de}{d\xi_1}(\xi_1)\,d\xi_1\right)\,,
\ee
This is the sum of an ``instantaneous'' and a ``relaxing'' contribution. If the identifications $q=X$ and $e=\Delta T$ are made, Eq. (\ref{nlfd2d}) is the Narayanaswamy expression Eq. (\ref{Nar}). This one cannot do {\it a priori}, however, because these variables generally belong to different energy bonds, but the extension needed to derive Eq. (\ref{Nar}) from Eq. (\ref{nlfd2d}) is straightforward, as we now proceed to show.

Assuming that the effort $e$ is the temperature variation $\Delta T$ -- the usual aging situation -- the corresponding flow variable is the entropy current, which is basically the heat current into the system \cite{meixner}. Equation (\ref{nlfd1}) generalizes straightaway to any number $n$ of energy bonds \cite{boc81}. The resulting equation involves a sum over all combinations of energy-bond indices, $j_1, ..., j_k$, such that for $i=1, ..., k$ each flow variable $f_{j_i}(t_i)$ in the right-hand-side cumulant of Eq. (\ref{nlfd1})'s generalization is paired to the same-index effort variable $e_{j_i}(t_i)$. If there are $n$ energy bonds, the leading term in the efforts is the following sum generalizing Eq. (\ref{nlfd1b}) 

\be\label{nlfd1b'}
\langle f(t_2)\rangle
\cong\beta_0\sum_{j=1}^n
\int_{-\infty}^{t_2} \left\langle f(t_2)f_j(t_1)\right\rangle e_j(t_1)dt_1\,.
\ee
If one is interested in how the quantity $X$ responds to a temperature variation, only one term in the above sum is of interest. This is the term coupling the $X$ energy bond to the thermal energy bond whose generalized charge -- the  time-integrated entropy current -- will be denoted by $q_S$. In this way one arrives at an expression of the form Eq. (\ref{nlfd2d}) with a different function $F(\xi)$, but the same material time, thus affirmatively answering question 3 in Sec. \ref{intro}. The function $F$ is given by $F(\xi_2 -\xi_1)=\langle q_S(\xi_1)X(\xi_2)\rangle$. Note that the above used identity $F'(0)=0$ applies also in the general case because it follows from equilibrium time reversibility, $\langle q_S(t_1)X(t_2)\rangle_0=\langle X(t_1)q_S(t_2)\rangle_0$.

\section{Identification of the material time}\label{clock}

We proceed to show that there is only one possible material time $\xi(t)$ obeying the MTTI requirement. Consider the configuration-space path of an aging system, $\bR(t)$. In the thermodynamic limit the relative fluctuations of distances go to zero. Because of this the ensemble-average symbols $\langle ... \rangle$ on the left-hand side of Eq. (\ref{MTTR2}) may be removed, leading to the following formulation of MTTI for the displacement

\be\label{R12eq}
R^2_{12}
\,=\, G(\xi(t_2)-\xi(t_1))\,,
\ee
where we have introduced the notation 

\be\label{R2not}
R^2_{12}\equiv \frac{\big(\bR(t_2)-\bR(t_1)\big)^2}{N}\,.
\ee
The division by the number of particles $N$ ensures a distance measure that is well defined in the thermodynamic limit ($N\rightarrow\infty$). 

First, uniqueness of the material time is demonstrated. Equation (\ref{R12eq}) applies, in particular, in equilibrium where $\gamma(t)$ is a constant, implying $\xi(t)= K t+{\rm Const.}$ In thermal equilibrium, the generic behavior is that $R^2_{12}$ is sublinear at short times, e.g., $R^2_{12}\sim (t_2-t_1)^x$ with $0<x<1$, and linear in $t_2-t_1$ at long times corresponding to ordinary diffusion \cite{sch00,dyr95,sch08}. The transition between the two regimes takes place around the system's average (alpha) relaxation time \cite{sch00,dyr06}, but we are not here interested in these details \cite{dyr95}. The important thing is that the equilibrium mean-square displacement at long times is always proportional to time. From this and the fact that $\xi=Kt+{\rm Const.}$ in equilibrium one concludes that 

\be\label{Geq}
G(\xi)\,\propto\,\xi\,\,\,(\xi\rightarrow\infty)\,.
\ee
This means that, if $\bR_0$ is a reference configuration far back in time on the aging system's trajectory, except for a multiplicative and an additive constant the only possible material time is defined by the squared distance from $\bR_0$ to $\bR(t)$ denoted by $R_{0t}$,

\be\label{mattimedef}
\xi(t)\equiv R^2_{0t}\,,
\ee
If a dimensionless material time is wanted, one may multiply $\xi(t)$ with the particle number density to the power $2/3$, but for simplicity we stick here to the above definition. 

Having established uniqueness of the material time, we proceed to show that the definition Eq. (\ref{mattimedef}) is consistent. For this to be the case the difference in material time between two events on the aging system's path in configuration space, $\xi(t_2)-\xi(t_1)$, must be independent of the choice of reference configuration. To show this we compare $R^2_{02}-R^2_{01}$ with the analogous quantity using a different reference configuration, $\bR(t'_0)$. Since $\bR(t_1)-\bR(t'_{0})=\bR(t_1)-\bR(t_{0})+\Delta\bR$ in which $\Delta\bR\equiv \bR(t_{0})-\bR(t'_{0})$, one has

\be
R^2_{0'1}
=R^2_{01}+\Delta\bR^2/N+2\big(\bR(t_1)-\bR(t_{0})\big)\cdot \Delta\bR/N\,.
\ee
Likewise

\be
R^2_{0'2}
=R^2_{02}+\Delta\bR^2/N+2\big(\bR(t_2)-\bR(t_{0})\big)\cdot \Delta\bR/N\,.
\ee
These equations lead to

\be\label{rconsist}
R^2_{0'2}-R^2_{0'1}
=R^2_{02}-R^2_{01}+2\big(\bR(t_2)-\bR(t_1)\big)\cdot \Delta\bR/N\,.
\ee
When $t_0$ and $t'_{0}$ are both far back in time, $\bR(t_2)-\bR(t_1)$ is uncorrelated with $\Delta\bR$ because in terms of the velocity $\bV\equiv \dot\bR$ one has
$\left[\bR(t_2)-\bR(t_1)\right]\cdot\Delta\bR=\int_{t_1}^{t_2}ds_1\int_{t_0}^{t'_{0}} ds_2 \bV(s_1)\cdot\bV(s_2)$, which goes to zero for $t_0\rightarrow -\infty$ and $t_{0'}\rightarrow -\infty$ since velocities far apart in time are uncorrelated (even for an aging system). Thus the last term of Eq. (\ref{rconsist}) vanishes in these limits, ensuring consistency of the material time definition Eq. (\ref{mattimedef}). Note that the reference configuration need not, in fact, be one of the system's distant past; any configuration far away may be selected as reference configuration. Note also that by differentiation of Eq. (24) one finds the following expression for the clock rate of Eq. (1): $\gamma(t)=2\int_{-\infty}^t {\bf V}(t')\cdot{\bf V}(t)dt'/N$.

\section{Two consequences}\label{conseq}

This section discusses two consequences of the formalism developed.

\subsection{The ``unique-triangles property''}

The configuration-space path of a system in thermal equilibrium has an interesting geometric property. Consider three times, $t_1< t_2< t_3$. Following Eq. (\ref{R2not}) the corresponding distances between the configurations on the system's path, $\bR(t_1)$, $\bR(t_2)$, and $\bR(t_3)$, are denoted by $R_{12}$, $R_{13}$, and $R_{23}$. If one defines $R_{\rm eq}(t)$ as the distance that the equilibrium system travels over time $t$, i.e.,

\be\label{Req}
R_{\rm eq}(t)\equiv \sqrt{\frac{\langle\Delta\bR^2(t)\rangle}{N}}\,,
\ee
we have $R_{12}=R_{\rm eq}(t_2-t_1)$, $R_{13}=R_{\rm eq}(t_3-t_1)$, and $R_{23}=R_{\rm eq}(t_3-t_2)$. Since $R_{\rm eq}(t)$ is an increasing function of $t$, this implies that the triangle formed by the points $\bR(t_1)$, $\bR(t_2)$, and $\bR(t_3)$ is unique in the following sense: If two of the triangle's side lengths are known, the third one is also known. For instance, if $R_{12}$ and $R_{23}$ are known, $R_{13}$ is determined since $R_{12}$ gives $t_2-t_1$ and $R_{23}$ gives $t_3-t_2$, from which $t_3-t_1=(t_3-t_2)+(t_2-t_1)$ and thus $R_{13}$ may be deduced. A system for which  any three points on its trajectory  determine a unique triangle in the above sense will be referred to as obeying the ``unique-triangles property''. 

It follows from Eq. (\ref{mattimedef}) that the unique-triangles property also applies for an aging system. This is because the difference of two material times, $\xi(t_2)-\xi(t_1)$, determines the distance between the corresponding configurations $\bR(t_1)$ and $\bR(t_2)$, $R_{12}$, and {\it vice versa}; likewise $\xi(t_3)-\xi(t_2)$ determines $R_{23}$. Thus $\xi(t_3)-\xi(t_1)=(\xi(t_3)-\xi(t_2))+(\xi(t_2)-\xi(t_1))$ on the one hand determines $R_{13}$ and on the other hand is uniquely determined by $R_{12}$ and $R_{23}$. -- Note the close similarity to the triangular relation Eq. (\ref{triang}) \cite{cug94}.

\subsection{A geometric interpretation of time-autocorrelation functions}

It is a property of equilibrium dynamics that knowledge of the value of one time-autocorrelation function determines the time difference in question and thus all other time-autocorrelation functions. Via MTTI as expressed in Eqs. (\ref{MTTdef}) and (\ref{MTTR2}) this property generalizes to aging systems; in particular it means that the value of $\langle A(t_1)A(t_2)\rangle$, which is a function of $\xi(t_2)-\xi(t_1)$, is in a one-to-one correspondence with the distance travelled, $R_{12}$. Thus one can define a {\it geometric} autocorrelation function as the average of $A(t_1)A(t_2)$ for all pairs of times of the aging system with the same distance $R_{12}$, corresponding to the same difference in material time. We denote this geometric autocorrelation function by $\langle A(0)A(R)\rangle$. MTTI leads to the identity

\be\label{aar}
\langle A(t_1)A(t_2)\rangle\,=\,\langle A(0)A(R)\rangle \bigg|_{R=R_{12}}\,.
\ee
Since this applies also when the system stops aging and is very close to equilibrium, the geometric autocorrelation function $\langle A(0)A(R)\rangle$ is identical to that characterizing  equilibrium (Appendix).

Equation (\ref{aar}) generalizes a proposal for a geometric interpretation of equilibrium time-autocorrelation functions put forward some time ago \cite{dyr97}. It was never published in a regular journal and is therefore briefly summarized in the Appendix. The idea is that the reason any time-autocorrelation function is small for large time separations is that the two relevant configurations are far from each other (we assume that $A$ only depends on the system's spatial coordinates, not the momenta). In glass-forming liquids all auto-correlation functions slow down in the same way as temperature is decreased, and moreover time-temperature superposition is often obeyed \cite{jak12}. These facts are easily understood from the geometric interpretation of time-autocorrelation functions (Appendix), according to which the slowing down of $R_{\rm eq}(t)$ upon cooling controls all time-autocorrelation functions in the same way, compare Eq. (\ref{aar}). TTS arises if the geometric autocorrelation function is independent of temperature, which seems to be a reasonable assumption.

\section{Possible numerical tests of the proposed framework}\label{tests}

The MTTI assumption can be checked by computer simulations. To do this it is first necessary to identify a system that obeys the Narayanaswamy formalism. This may be challenging because computers are not yet able to simulate realistic aging situations, so temperature jumps larger than a few percent may be needed in simulations \cite{vol13}. 

Once a suitable model system has been identified, the following tests can be performed: 

\begin{enumerate}

\item {\it Predicting nonlinear aging from linear aging.} The ultimate test of Eq. (\ref{Nar}) is to investigate whether information from very small temperature-jump simulations, i.e., aging in the linear regime, is enough to predict aging following larger temperature jumps when the material time is defined by Eq. (\ref{mattimedef}).

\item {\it The unique-triangles property for an aging system.} 
This can be investigated in the same way the triangular relation is checked \cite{cha07}: First the system's path in configuration space $\bR(t)$ is traced out and stored. Different triplets of configurations, $\bR(t_1)$, $\bR(t_2)$, and $\bR(t_3)$, are then picked out and the corresponding distances $R_{12}$, $R_{13}$, and $R_{23}$ evaluated. The system obeys the unique-triangles property if for a given value of $R_{12}$ there is a one-to-one correspondence between  $R_{13}$ and $R_{23}$. In practice, this may be checked by plotting $R_{13}$ versus $R_{23}$ for a narrow range of $R_{12}$ values.

\item {\it The geometric ansatz for the time-autocorrelation functions.}
To check Eq. (\ref{aar}) for an aging system one may proceed as follows. First evaluate the equilibrium geometric autocorrelation function $\langle A(0)A(R)\rangle_0$. This may be done by tracing out and storing the system's path in configuration space $\bR(t)$. For many pairs of configurations on this path, $\bR_1$ and $\bR_2$, the distance $R_{12}$ and product $A(\bR_1)A(\bR_2)$ is evaluated; $\langle A(0)A(R)\rangle_0$ is then the average of these products with $R=R_{12}$. After this, the same procedure is performed for an aging system to see whether the same geometric autocorrelation function applies. An equivalent test is to investigate whether $\Delta A_{12}\equiv A(\bR_2)-A(\bR_1)$ is the same function of the distance $R_{12}$ in equilibrium and during aging, which can be done by plotting $\Delta A_{12}$ versus $R_{12}$ and comparing the two plots.

\end{enumerate}

If simulations have also identified a model for which the Narayanaswamy theory does not apply, it would be interesting to see whether this system violates the unique-triangles property. This may illuminate whether that property is a good ``thermometer'' for which systems obey the Narayanaswamy theory and which do not.

\section{Concluding remarks}\label{end}

The purpose of this paper was to address the challenge of justifying from basic principles the phenomenological, but indisputably successful 44 years old Narayanaswamy description of aging. The above reasoning should not be regarded as an attempt to formulate a compelling theory for physical aging, but merely as suggesting one possible way to go about this subject.

Taking material-time translational invariance (MTTI) as the starting point, we have seen  that 

\begin{itemize}

\item MTTI is equivalent to the triangular relation discovered by Cugliandolo and Kurchan in 1994 from theoretical studies of spin glass aging \cite{cug94}.

\item MTTI in conjunction with the 1981 Bochkov-Kuzovlev exact nonlinear fluctuation-dissipation theorem leads to the Narayanaswamy aging theory for small temperature variations.

\item There is only one possible material time obeying MTTI, namely that defined by the distance squared to a configuration of the system's distant past.

\item MTTI implies a geometric picture of time-autocorrelation functions for an aging system, according to which the time-autocorrelation is regarded as a {\it spatial} autocorrelation evaluated at the distance travelled in configuration space.

\end{itemize}

An open question is the origin of the intriguing extension of Onsager reciprocity to aging systems' time autocorrelation functions, i.e., the finding that in certain spin glass models $\langle A(t_1)B(t_2)\rangle=\langle B(t_1)A(t_2)\rangle$ as discussed by Cugliandolo and Kurchan \cite{cug99} and by Franz and Virasoro \cite{fra00}. This applies trivially in equilibrium because of time-reversal invariance. Whenever Onsager reciprocity applies for an aging system, it indicates an element of reversibility during aging \cite{cug99,fra00,fra15}. Given the analogous aging behavior of spin glasses and real glasses, it seems likely that Onsager reciprocity may also apply for the latter, meaning that aging for relatively small temperature variations has an element of reversibility. This is consistent with the fact that the algebraic operation $*$ discussed in Sec. \ref{triangsec} is commutative.

In Sec. \ref{intro} we listed four questions relating to the Narayanaswamy formalism. The above developments addressed the first three of these. Thus we have shown 1) how, by use of the Bochkov-Kuzovlev exact nonlinear fluctuation-dissipation theorem and MTTI, the highly nonlinear aging phenomenon may be reduced to a linear material-time convolution integral; 2) the material time was identified as the distance to a reference configuration far away; 3) whether all quantities age following the same material time is answered with a yes since the material-time definition of Eq. (\ref{mattimedef}) is unique. 

The remaining question from Sec. \ref{intro} is: what controls the material time? The traditional answer to this is the fictive temperature $T_f$ \cite{nar71,too46}, which by definition quantifies the structure in such a way that in $T_f=T$ in equilibrium. The fictive-temperature concept is hand waving, however, and it would be nice to have a microscopic understanding of what controls the material time's clock rate $\gamma(t)$ of Eq. (\ref{eq1}) \cite{scherer,ols98,hod95}. This amounts to solving one of the deepest problems in glass science: what controls the equilibrium relaxation time's temperature dependence \cite{dyr06}. Aging studies may contribute to solving this problem by providing information beyond that obtainable from linear-response experiments. For instance, an aging experiment could in principle determine whether following a temperature jump there is an instantaneous change of the activation energy of the clock rate $\gamma(t)$ as predicted, e.g., by the shoving model \cite{dyr06,ols98,dyr96}, or whether there is no such instantaneous change, which is the expectation, e.g., from the Adam-Gibbs configurational-entropy model \cite{ada65}. Another example of aging studies with consequences for the general understanding of viscous liquid dynamics is the suggestion that the potential energy controls the material time and thus, in particular, glass-forming liquids' equilibrium average relaxation time \cite{ado07}. This interesting idea connects to and includes the material time of rheology \cite{sol98,fie00}, which was Narayanaswamy's source of inspiration \cite{nar71}.

\begin{acknowledgments}
The author is indebted to Jorge Kurchan for a very useful discussion. The center for viscous liquid dynamics ``Glass and Time'' is sponsored by the Danish National Research Foundation via grant DNRF61.
\end{acknowledgments}

\section*{Appendix: Geometric theory of equilibrium time-autocorrelation functions}

As in the main paper, the system's path in configuration space as a function of time is denoted by $\bR(t)$ and we define the distance $R_{12}$ between two configurations $\bR(t_1)$ and $\bR(t_2)$ by Eq. (\ref{R2not}). As mentioned, fluctuations become insignificant in the $N\rightarrow\infty$ limit, and consequently the distance between the system's configurations at two times is a unique number. This simplification applies in  equilibrium, as well as for an aging system. In thermal equilibrium the displacement of the system during time $t$ is denoted by $R_{\rm eq}(t)$ (Eq. (\ref{Req})). Note that this quantity is experimentally accessible via the incoherent intermediate scattering function. 

Some time ago I proposed a geometric view of thermal-equilibrium time-autocorrelation functions \cite{dyr97}. The background was the facts that time-temperature superposition (TTS) often applies for linear-response functions of glass-forming liquids and that, while the average relaxation time $\tau(T)$ is always strongly temperature dependent, $\tau(T)$ of different linear-response functions usually varies with temperature in exactly the same way \cite{jak12}. A simple way to understand these facts is that there is a temperature-independent {\it geometric autocorrelation function},  

\be
\langle A(0)A(R)\rangle\,.
\ee
by which is meant the quantity defined by averaging over all pairs of points along the system's path in time separated by the distance $R=R_{\rm eq}(t)$, corresponding to the time interval $t=t_2-t_1$ between two configurations, $\bR(t_1)$ and $\bR(t_2)$. This distance is unique for given $t$. $R_{\rm eq}(t)$ is an increasing function of $t$ and thus in a one-to-one correspondence with $t$. In terms of the geometric autocorrelation function an equilibrium time-autocorrelation function is given \cite{dyr97} by 

\be\label{geomans}
\langle A(0)A(t)\rangle_0
\,=\,\langle A(0)A(R)\rangle \bigg|_{R=R_{\rm eq}(t)}\,.
\ee
As it stands, Eq.(\ref{geomans}) is a tautology because the geometric average $\langle A(0)A(R)\rangle$ is {\it defined} to make Eq. (\ref{geomans}) apply at the thermodynamic state point in question. 

One of the consequences of Ref. \onlinecite{dyr97} is that TTS finds a natural explanation if what happens when temperature is changed is simply the following: the geometric autocorrelation function $\langle A(0)A(R)\rangle$ is unchanged, and $R_{\rm eq}(t)$ is also unchanged except for an overall scaling of time reflecting the slowing down upon cooling. This geometric ``explanation'' of TTS admittedly presupposes TTS for the mean-square displacement, but there is considerable evidence for this from theoretical and numerical studies of hopping in highly disordered landscapes \cite{dyr95,sch00,sch08}. 

In Ref. \onlinecite{dyr97} we proposed an ansatz for calculating the geometric autocorrelation function $\langle A(0)A(R)\rangle$ via a ``double-canonical'' statistical-mechanical average, and the theory was validated by simulations of a one-dimensional double-potential model. 

A simple example of Eq. (\ref{geomans}) is the case of a Gaussian geometric autocorrelation function, $\langle A(0)A(R)\rangle\propto\exp(-R^2/2R_0^2)$, and linear diffusion, $R_{\rm eq}^2(t) =6D t$, where $D$ is the diffusion constant. This leads to an exponentially decaying time-autocorrelation function, the so-called Debye relaxation that is the simplest linear-response situation. A more realistic case also assumes a Gaussian geometric autocorrelation function, but combines this with the fact that the short-time mean-square displacement is usually subdiffusive \cite{sch00}. This leads to a high-frequency (short-time) non-Debye behavior. In experiments the relaxation is often Debye-like on the low-frequency side of the loss peak \cite{nie09} -- at least for non-polymeric systems -- which implies that a Gaussian decay of the geometric autocorrelation function {\it must} apply at long distances. 

The assumption that the equilibrium geometric autocorrelation function $\langle A(0)A(R)\rangle$ is temperature independent is a quantitative expression of the physical idea that at different temperatures the system paths are very similar, except for the fact that the rate of motion has changed \cite{par09}. Whenever this applies, it is natural to expect that the same geometric autocorrelation function describes a system aging at these temperatures.

\end{document}